\documentclass[prl,showpacs,twocolumn,superscriptaddress,aps]{revtex4-1}
\usepackage[utf8]{inputenc}
\usepackage[american,british]{babel}
\usepackage[T1]{fontenc}
\usepackage[pdftex]{graphicx}  
\usepackage{graphicx, xcolor}
\usepackage{dcolumn}
\usepackage{bm}
\usepackage{amsmath,amsthm,amssymb}
\usepackage{amsmath}
\usepackage{color}
\usepackage{verbatim}
\usepackage{ulem}

\definecolor{darkGreen}{RGB}{0,110,0}
\definecolor{darkBlue}{RGB}{0,0,130}
\usepackage[colorlinks,citecolor=darkGreen,linkcolor=darkBlue,urlcolor=blue,hyperindex]{hyperref}

\newcommand{\ket}[1]{\left| #1 \right\rangle}

\begin{document}
\title{Many-body localization dynamics from gauge invariance}
\author{Marlon Brenes}
\affiliation{The Abdus Salam International Center for Theoretical Physics, Strada  Costiera  11,  34151  Trieste,  Italy}
\author{Marcello Dalmonte}
\email{Corresponding author: mdalmont@ictp.it}
\affiliation{The Abdus Salam International Center for Theoretical Physics, Strada  Costiera  11,  34151  Trieste,  Italy}\author{Markus Heyl}
\affiliation{Max Planck Institute for the Physics of Complex Systems, Dresden 01187, Germany}
\author{Antonello Scardicchio}
\affiliation{The Abdus Salam International Center for Theoretical Physics, Strada  Costiera  11,  34151  Trieste,  Italy}\affiliation{INFN Sezione di Trieste, Via Valerio 2, 34127 Trieste, Italy}

\date{\today}


\begin{abstract}
We show how lattice gauge theories can display many-body localization dynamics in the absence of disorder. Our starting point is the observation that, for some generic translationally invariant states, Gauss law effectively induces a dynamics which can be described as a disorder average over gauge super-selection sectors. We carry out extensive exact simulations on the real-time dynamics of a lattice Schwinger model, describing the coupling between U(1) gauge fields and staggered fermions. Our results show how memory effects and slow, double-logarithmic entanglement growth are present in a broad regime of parameters - in particular, for sufficiently {\it large} interactions. These findings are immediately relevant to cold atoms and trapped ions experiments realizing dynamical gauge fields, and suggest a new and universal link between confinement and entanglement dynamics in the many-body localized phase of lattice models. 
\end{abstract} 

\maketitle

\paragraph{Introduction. -- } 

Over the last two decades, the impressive developments in harnessing matter at the single quantum level have paved the way to the investigation of real-time dynamics in controlled quantum systems with an unparalleled degree of accuracy~\cite{bloch2012quantum,blatt2012quantum,Georgescu2014}. These progresses, spanning as diverse fields as cold atoms in optical lattices, trapped ions, superconducting circuits, and more, have reinvigorated the theoretical interest in the dynamics of closed quantum systems~\cite{Polkovnikov2011b}. A paradigmatic example in this direction is the quest for generic systems in which the interplay of disorder and interactions prevents thermalization, a scenario dubbed many-body localization (MBL) \cite{Basko:2006hh}. In this new dynamical phase, the discreteness of the local observables spectra, typical of quantum mechanics, endows the system with a set of local integrals of motion which freeze transport and localize excitations\cite{vznidarivc2008many,pal2010MBL,bardarson2012unbounded,huse2013phenomenology,serbyn2013local,ros2015integrals,Imbrie2016,Imbrie2016c} (for reviews see\cite{nandkishore2015many,Altman2015,imbrie2017review}). While this phenomenon has been predicted, and signatures observed in numerics and experiments for a variety of model Hamiltonians such as Hubbard models and spin chains~\cite{nandkishore2015many,Bloch2015,monroe2015}, it is an open question to which extent such lack of thermalization can occur in fundamental theories of matter, and in particular, if gauge invariance can play a role in the mechanism.

In this paper we show the emergence of MBL dynamics in Lattice Gauge Theories (LGTs)~\cite{wilson1974confinement,kogut1975hamiltonian,Montvay1994} in the absence of any disorder. Our work is immediately motivated by recent theoretical proposals~\cite{Banerjee2013,Tagliacozzo2013,Zohar:2013kb,Notarnicola2015a,Kasper:2016aa,Muschik:2016aa} and experimental demonstration~\cite{martinez2016real} of LGT dynamics in synthetic quantum systems, and by the paradigmatic importance played by gauge theories. The latter describe a plethora of physical phenomena, from fundamental interactions in particle physics~\cite{Montvay1994} to the low-energy dynamics of frustrated quantum magnets~\cite{Lacroix2010}, and are instrumental in designing quantum computing architectures which show inherent protection against noise \cite{kitaev2010topological}. As such, addressing their real-time dynamics is of profound interest from a variety of perspectives, regarding both the basic understanding of lattice field theories, and the possibility of safely storing quantum information via localization in quantum memories, further boosting their resilience.

\begin{figure}
\includegraphics[width=0.99\columnwidth]{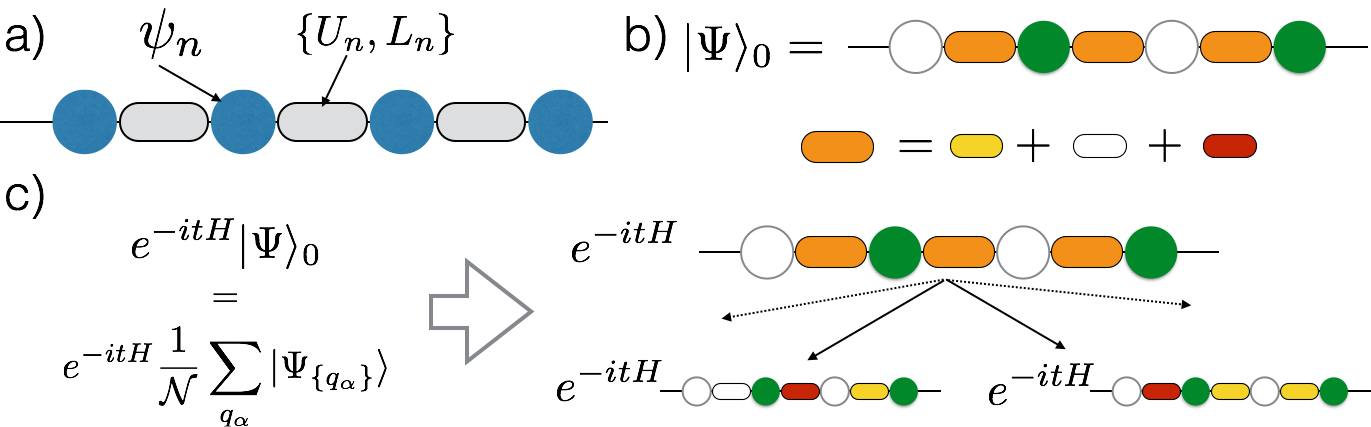}
\centering
\caption{Panel (a): schematic of a (1+1)-d lattice gauge theories, with matter fields $\psi_n$ defined on the vertices, and pairs of conjugate variables $\{U_n, L_n\}$ defined on bonds. (b) Typical initial states used in the simulations: fermions are arranged in an alternate pattern of empty and occupied sites (corresponding the the bare vacuum), while gauge fields are in an equal weight superposition of electric field eigenstates with eigenvalues $(0, \pm1)$. (c) Schematics of the real time dynamics, which starting from $|\Psi\rangle_0$, can be decomposed in $\mathcal{N}$ superselection sector (see text). }
\label{fig:1}
\end{figure} 

Our main finding is that, starting from translational invariant states, the dynamics of LGT is profoundly influenced by the presence of super-selection sectors - a key element that stems directly from gauge invariance (see Fig.~\ref{fig:1}). Even though MBL in systems without disorder has already been discussed and debated over the last years~\cite{Schiulaz2015,Hickey2016,Horssen2015,Papic2015,schiulaz2014glass,pino2015metallic,yao2016quasi,Prem:2017aa,Kim:2016aa}, the presence of such super-selection sectors provides a pristine mechanism for localization dynamics, whose origin can be conveniently tracked by an exact integration of the gauge fields in (1+1)-d lattice gauge theories describing matter coupled to gauge fields. 

Following this analytical understanding, we provide extensive numerical evidence for MBL dynamics in the lattice Schwinger model - a (1+1)-d version of quantum electrodynamics, with U(1) gauge fields coupled to Kogut-Susskind fermions~\cite{kogut1975hamiltonian,Montvay1994}- based on both strong memory effects and entanglement dynamics. The latter is characterized by a sub-logarithmic growth of the bipartite entanglement entropy, which we interpret as a transport-inhibiting mechanism due to confinement. Our results have immediate experimental relevance in trapped ions systems, where the Schwinger model has already been experimentally realized~\cite{martinez2016real}, and in cold atom gases in optical lattices, where various implementations schemes for U(1) LGT~\cite{Banerjee2013,Tagliacozzo2013,Zohar:2013kb,Notarnicola2015a,Kasper:2016aa,Muschik:2016aa} have been put forward.

\paragraph{Model Hamiltonian. --} We are interested here in the dynamics of Abelian lattice gauge theories on a one-dimensional lattice. While our approach is general and can be applied to arbitrary Abelian (and continuos non-Abelian) gauge theories in both Wilson~\cite{kogut1975hamiltonian,Montvay1994} and quantum link formulations~\cite{Chandrasekharan1997}, for the sake of simplicity, we focus in the following on the $U(1)$ Wilson LGT, the lattice Schwinger model (SM). Despite its simplicity, the SM still displays many paradigmatic features also found in more complex gauge theories, such as confinement and string breaking dynamics. The system is described by a Kogut-Susskind Hamiltonian \cite{kogut1975hamiltonian} of the form:
\begin{eqnarray}\label{Eq_LatticeHamiltonian}
H&=&-i w\sum_{n=1}^{N-1}\left[{\psi}^{\dag}_nU_{n}{\psi}_{n+1}-\textrm{h.c.}\right]\nonumber\\
&+&J\sum_{n=1}^{N-1} {L}_n^2+m \sum_{n=1}^{N}(-1)^n {\psi}^{\dag}_n{\psi}_n ,\ \  \  \ 
\end{eqnarray}
where ${\psi}_n$ are fermionic annihilation operators defined on vertices, $U_{n} = e^{i\varphi_{n}}$ are $U(1)$ parallel transporters defined on bond $(n,n+1)$, whose corresponding electric field operator is defined as $L_n=-i\partial/\partial\varphi_{n}$, so that $[L_{n},U_{n} ] = U_{n}$ which ensures gauge covariance. The first term describes the gauge-matter coupling; the second is a model dependent electric field contribution~\cite{Hamer1997,Muschik:2016aa}, and indicates the strength of the inter-particle interaction mediated by the gauge field; finally, the last term describes the staggered mass of the fermions (which we will set to 0 in the following). While we are not interested in the continuum limit of the theory here, it is possible to safely take it using the lattice formulation and properly scaling the coupling parameters~\cite{Hamer1997,Muschik:2016aa}. As in conventional lattice gauge theories, gauge invariance is manifest after defining a set of generators, $G_n=L_n-L_{n-1} -\psi^\dagger_n\psi_n+\frac{1}{2}\left[1-(-1)^n\right]$, which satisfy $[H, G_n]=0$. States in the Hilbert space as defined by Gauss law, which reads
\begin{equation}
G_n|\Psi_{\{q_\alpha\}}\rangle = q_n  |\Psi_{\{q_\alpha\}}\rangle \,,
\end{equation}
where $\{q_\alpha\}$ represents the distribution of background charges $q_n$ on each state in the Hilbert space, defining a superselection sector. In a U(1) LGT, $q_n\in [-\infty, \infty]$. We emphasize that the presence of Gauss law and superselection sectors has nothing to do with integrability, but stems solely from gauge invariance.

\paragraph{Superselection sectors as a mechanism for disorder-free localization. -}The presence of these superselection sectors drastically affects the system dynamics. In (1+1)-d, this can be elegantly seen by integrating out the gauge fields, and then studying the resulting Hamiltonian acting on the matter degrees of freedom. Below we show that this procedure indicates that MBL is a rather generic scenario for LGT in the absence of any underlying disorder - e.g., systems with homogeneous couplings and homogeneous initial states show robust memory effects and slow (sub-logarithmic) growth of entanglement entropy as a function of time.

We investigate the time evolution of initial states of the form:
\begin{equation}
|\Psi\rangle_{0} = |0101...\rangle_\psi\otimes |\bar{L}_n...\rangle_\sigma \, ,
\end{equation}
where the fermions are in a Neel state (corresponding to the bare vacuum of staggered fermions), and the gauge fields $L_n$ are in an equal weight superposition of $\{-1, 0, 1\}$ at each site. This state is translational invariant up to translations of two lattice spacings, and can be decomposed into U(1) superselection sectors as:
\begin{equation}\label{psi_in}
|\Psi\rangle_{0} =\frac{1}{\mathcal{N}^{1/2}} \sum_{\bar{q}_n=0, \pm1}|0101..\rangle_\psi\otimes |\bar{q}_n...\rangle_\sigma = \frac{1}{\mathcal{N}^{1/2}} \sum_{q_\alpha}|\Psi_{\{q_\alpha\}}\rangle \, .
\end{equation}
with $\mathcal{N}$ denoting the total number of different superselection sectors. In order to derive the dynamics within each sector, we analytically integrate out the gauge fields~\cite{Hamer1997}. We assume $\bar{L}_{0}=0$ to minimize boundary effects, and apply a Jordan-Wigner transformations to the fermionic fields in order to re-cast the dynamics as a spin model, $\psi^\dagger_n \psi_n= (\sigma^z_n+1)/2 $. The gauge fields can be sequentially integrated out by noting that 
\begin{equation}
L_{\ell} = L_{\ell-1} + (\sigma^z_\ell +(-1)^\ell)/2 + q_\ell \, ,
\end{equation}
which simply describes the fact that, given the value of the ingoing electric field on the left side of a site, and given the values of the dynamical and static charge, the value of the outgoing electric field is unambiguously fixed.

After integration, the resulting Hamiltonian dynamics crucially depends on $\{q_\alpha\}$ - that is, states in different superselection sectors will evolve according to different Hamiltonians $H^{\{q_\alpha\}}$. This is a direct consequence of the fact that, because of Gauss law, different superselection sectors describe dynamics subject to different static charge configurations. In each sector, the corresponding Hamiltonian is made of two contributions, $H^{\{q_\alpha\}} = {H}_{\pm}+H_{In}^{\{q_\alpha\}}$. The first one describes electron-gauge coupling, which is not sensitive to background charges, and is given by ${H}_{\pm}=w\sum_{n=1}^{N-1}\left[{\sigma}_n^{+}{\sigma}_{n+1}^{-}+\textrm{h.c.}\right]$. The second term originates from the electric field potential term, which is now a function of the fermionic populations only. It contains an interaction part:
\begin{eqnarray}
{H}_{ZZ}\!&=&\!\frac{J}{2}\sum_{n=1}^{N-2}\sum_{l=n+1}^{N-1}(N-l){\sigma}_n^{z} {\sigma}_l^z
\end{eqnarray}
which is related to the linear growth of Coulomb interactions in one-dimensional systems, and single spin terms:
\begin{eqnarray}
{H}_{Z}^{\{q_\alpha\}}\!&=
&\!\frac{J}{2}\sum_{n=1}^{N-1} (\sum_{\ell=1}^n \sigma^z_\ell) \left[(\sum_{j=1}^n q_j) - n\text{mod}2 \right].
\label{eq:HamZ}
\end{eqnarray}
Crucially, this last part of the Hamiltonian depends explicitly on the superselection sector via $\{q_\ell\}$. As such, starting from initial states of the form in Eq.~\eqref{psi_in}, the system dynamics is dictated by a {\it charge distribution average}, that is:
\begin{equation}\label{H_q}
|\Psi(t)\rangle = e^{-itH}|\Psi\rangle_0 = \frac{1}{\mathcal{N}^{1/2}}\sum_{\{q_\alpha\}} e^{-itH^{\{q_\alpha\}}}|\Psi_{\{q_\alpha\}}\rangle \, ,
\end{equation}
which is effectively describing a disorder average, since the terms in ${H}_{Z}^{\{q_\alpha\}}$ effectively act as a (correlated) disorder for the spin dynamics. The observation in Eq.~\eqref{H_q} is applicable to arbitrary Abelian Wilson theories, and even non-Abelian QLM, in one-dimensional systems. For the specific case of $\mathbb{Z}_2$ quantum link models, our theory (see Ref.~\cite{supmat}) recovers the results of Ref.~\cite{smith2017disorder}, which reported Anderson localization in disorder free models. The same reasoning can be applied to two-dimensional systems, in particular, to the evolution of quenched gauge theories. It explicitly shows that, contrary to conventional spin models, the dynamics of systems endowed by a gauge symmetry can naturally lead to phenomena related to disordered systems, even in the absence of any disorder (both on the initial state, and in the dynamics). This happens for generic initial states which are in product form of the matter and gauge fields, as in those cases, the weight of superselection sectors with effectively no disorder decreases exponentially with system size. Finally, we emphasize that the mechanism discussed here is very different from the disorder-free localization dynamics discussed in the context of fractons~\cite{Prem:2017aa,Kim:2016aa}, which relies on slow dynamics of topological excitations and separation of energy scales~\footnote{Note that the notation of superselector sector has a different meaning in our work and Ref.~\cite{Prem:2017aa}.}\footnote{The possibility that the constrained Hilbert space be the sole reason of breaking of ETH has been ruled out in \cite{chandran2016eigenstate}.}.

\begin{figure}
\centering
\includegraphics[width=1.01\columnwidth]{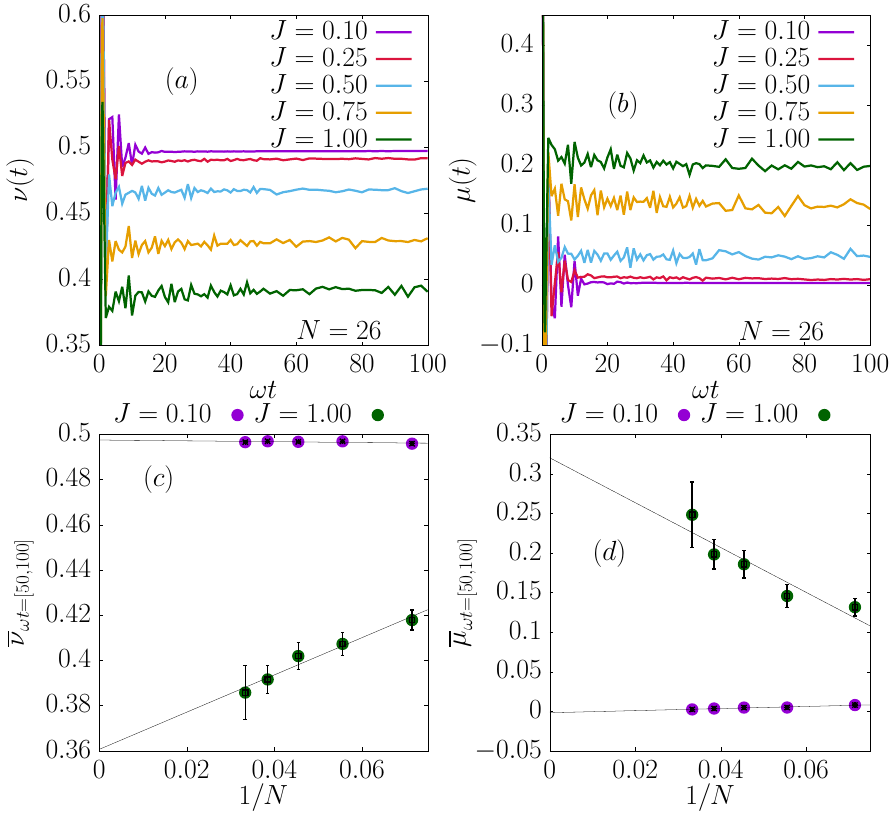}
\caption{(Color online) Dynamics of the overall staggered fermion occupation (a) and central staggered fermion occupation (b) from a bare vacuum initial state for different values of coupling strength parameter $J$, and $N = 26$. (c-d) Final time value of $\nu(t)$ and $\mu(t)$, respectively, averaged from $t=50$ to $t=100$ for different system sizes up to $N=30$ and two limiting values of $J$. Absolute error due to averaging realizations is shown as vertical bars for each point.}
\label{fig:2}
\end{figure} 

\paragraph{Many-body localization dynamics in U(1) lattice gauge theories. --} While our theory predicts a generic mechanism for effectively disordered dynamics in LGTs, the question if this finally leads to localization behavior has to be addressed by non-perturbative methods. We thus turn to a numerical investigation of the system dynamics described by Eq.~\eqref{H_q}. We address this by employing a computationally optimized approach to the method of Krylov subspaces - for details, see Ref.~\cite{supmat}. As a first figure of merit, we focus on the staggered occupation of the fermions:
\begin{eqnarray*}
\nu(t)=\frac{1}{2N}\sum_{n=1}^N  \langle  (-1)^n \psi_n^\dagger(t) \psi_n(t) +1\rangle,
\end{eqnarray*}
which, by a Jordan-Wigner transformation, can be expressed as $\nu(t)=\frac{1}{2N}\sum_{n=1}^N  \langle  (-1)^n \sigma^z_n +1\rangle$ by transforming fermionic fields to spin operators. Since we employ staggered fermions, this quantity corresponds to the total number of particles created in the system.  In addition, we also monitor the staggered population for the central two sites of the system:
 \begin{eqnarray*}
\mu(t)=\langle \hat{n}_{\frac{N}{2}}(t) - \hat{n}_{\frac{N}{2} + 1} \rangle,
\end{eqnarray*}
where $\hat{n}_{i}(t) \equiv \psi_i^\dagger(t)\psi_i(t)$ is the fermion counting operator on site $i$. 

In Fig.~\ref{fig:2}a-b, we plot typical results of our simulations for $N=26$. In the weak coupling limit, both quantities reach their average thermodynamic value: 0.5 and 0, respectively. In contrast, for the strong coupling phase, the system does not relax for both observables (we have carried out simulations up to times $10^{20}$ until $L=14$ to check this). This behavior is analyzed using finite-size scaling in Fig.~\ref{fig:2}c-d: for $J=0.1$, both quantities approach their thermodynamic value as $N$ is increased. Instead, for $J=1.0$, for both $\nu(t)$ and $\mu(t)$ the deviation from thermodynamic values actually {\it increases} as a function of system size, signaling strong memory effects. 

\begin{figure}[t]
\centering
\includegraphics[width=1.01\columnwidth]{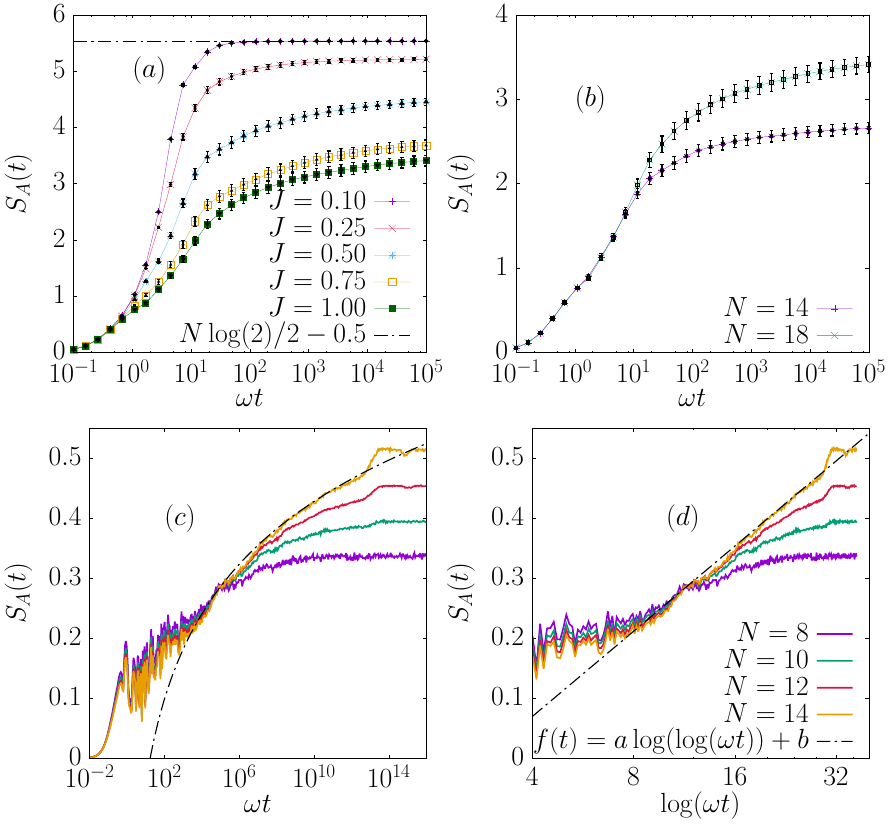}
\caption{(Color online) Time evolution of the bipartite von Neumann entanglement entropy $S_A(t)$ from a bare vacuum initial state. (a) $S_A(t)$ for different values of coupling strength parameter $J$ and $N=18$. The dashed line refers to the value $N\log(2)/2-1/2$ of the entanglement entropy as expected for a random state. (b) $S_A(t)$ on intermediate time scales for $J=1$ and different system sizes $N$. (c) Long-time growth of $S_A(t)$ for $J=10$ including a fit (solid line) $f(t)=a\log(\log(wt))+b$ to the $N=14$ data. (d) The same data is plotted on a double logarithmic scale.}
\label{fig:3}
\end{figure} 

Another key aspect of MBL is a very slow propagation of quantum information~\cite{vznidarivc2008many,pal2010MBL,bardarson2012unbounded,huse2013phenomenology}, which can be studied by monitoring the bipartite von Neumann entanglement entropy after the quench, defined as:
\begin{equation}
S_{A} = -\textrm{Tr}_A \rho_A\log \rho_A \, ,
\end{equation}
where $\rho_A$ denotes the reduced density matrix of the state in the region A, which in the following will be half system. For systems whose dynamics is captured by the eigenstate thermalization hypothesis, one expects that, after a quantum quench, the bipartite entanglement entropy grows linearly as a function of time; instead, MBL systems are characterized by slow entanglement spreading, which is typically logarithmic in time~\cite{DeChiara2006,vznidarivc2008many,bardarson2012unbounded}.

In Fig.~\ref{fig:3}a-b, we plot $S_{A}$ as a function of time for the same parameter regimes as in Fig.~\ref{fig:2}. While for small $J=0.1$ the entropy grows faster than logarithmically (and approaches a saturation value of order $N/2\log(2)-0.5$, as expected for a random state~\cite{Page1993}), in the localized phase the growth is extremely slow: in particular, it is slower than $\log(t)$. Using full diagonalization of the Hamiltonian, we further analyze this behavior in Fig.~\ref{fig:3}c-d for the strongly interacting regime, $J=10$: in this case, the decay is consistent with a double logarithmic behavior, as it is definitely slower than $(\log(t))^\alpha$. This type of growth has never been reported in MBL systems, where the typical $\log(t)$ behavior can be inferred in the local integral of motion picture and considering short-ranged interactions. 

In LGT however, we argue that confinement plays an important role in determining entanglement dynamics. For this purpose, we compare the present case, which is confining, with a deconfined regime. The latter can be achieved in the presence of a finite $\theta$-angle with $\theta=\pi$~\cite{coleman1976more}; in this regime, including a four-Fermi coupling interaction $Vn_jn_{j+1}$, the dynamics of the states $|\Psi\rangle_0$ is mapped exactly to an XXZ chain with (correlated) disorder in $\sigma^z$ terms, a model whose transport and entanglement properties have been extensively studied in the context of MBL.\cite{Pal2010,serbyn2013universal,de2013ergodicity,alet2015,John2015TotalCorrelations,Znidaric2016Diffusive}

The comparison between the two cases reveals that the fact that excitations are confined drastically changes entanglement spreading, further decreasing its growth from logarithmic to sub-logarithmic. The fact that confinement affects even this 'high-energy' behavior is not surprising, as it describes the behavior of large interparticle separations, very far from ground state physics.

\begin{figure}[t]
\centering
\includegraphics[width=1.01\columnwidth]{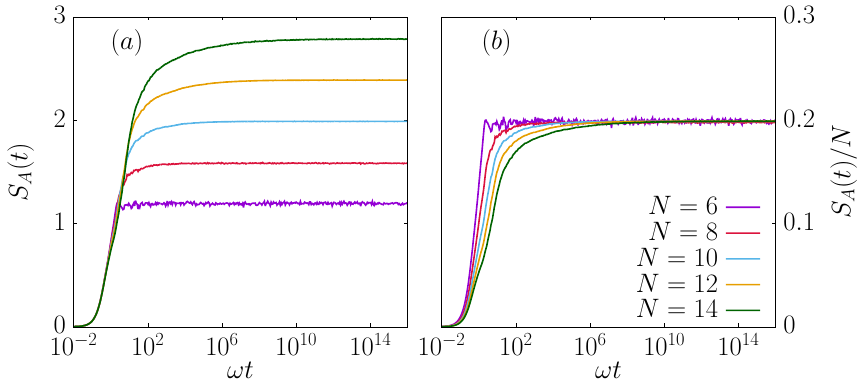}
\caption{(Color online) (a) Asymptotic long-time dynamics of the entanglement entropy density $S_A(t)$ for different system sizes $N$ at $J=1$ suggesting saturation to an extensive value by plotting the entanglement entropy density $S_A(t)/N$ (b).}
\label{fig:4}
\end{figure}

Finally, in Fig.~\ref{fig:4}, we show the long-time dynamics of $S_A$ and in particular its saturation value. As illustrated in Fig.~\ref{fig:4}b, the latter scales linearly as a function of system size, another characteristic feature of MBL dynamics.

\paragraph{Conclusions and outlook.} We have shown how many-body localization dynamics can naturally emerge in lattice gauge theories in the absence of any disorder. The physical interpretation is that, due to the presence of superselection sectors, the dynamics of translational invariant states is effectively described by an average over random charge configurations, with electrons interacting via long-range Coulomb potential. While our results are derived from a fully local field theory, after gauge field integration they signal that certain types of long-range interactions {\it per se} are not sufficient to delocalize the charges. This is surprising in view of earlier results \cite{PhysRevLett.64.547,burin2006energy,yao2014many} but in line with recent treatments of the long-range interactions in a non-perturbative way \cite{nandkishore2017amany}. We remark that there is a key difference between gauge theories and, say, generic long-range XY models: only the former can be exactly recast in a local fashion. It would be interesting to see whether matching this key requirement leads to a strict criterion for MBL in long-range systems.

Moreover, we provide numerical evidence that shows how confinement substantially favours localization, in particular, by inhibiting entanglement growth in a qualitative stronger manner with respect to the local integral of motions in statistical mechanics models of MBL. Our results point toward intriguing possibilities to study MBL dynamics in 2D LGT: in particular, due to the modest Hilbert space dimension growth in simple quantum link models, numerical simulations are expected to be comparatively easier with respect to spin models (especially for U(1) theories), and gauge invariant tensor network methods~\cite{Schollwock2011,Banuls,Rico2014a,Buyens2014a,Tagliacozzo2014} could be used thanks to the considerably slower entanglement growth in LGT.

\paragraph{Acknowledgements.}

We thank R. Fazio, J. Goold, R. Moessner, and R. Nandkishore for discussions. A.~S.\ acknowledges financial support from a Google Faculty Award. M.~B.\ and M.~D.\ thank Ivan Girotto for support, and acknowledge computing resources at CINECA through the Italian SuperComputing Resource Allocation - ISCRA grant LoLaGa.~M.~H.\ acknowledges support by the Deutsche Forschungsgemeinschaft via the Gottfried Wilhelm Leibniz Prize programme.

{\it Note added. -} While this manuscript was in preparation, a preprint appeared~\cite{Smith:2017aa} where a $\mathbb{Z}_2$ quantum link model with four-Fermi coupling was considered. Our theoretical analysis predicts disorder free MBL and logarithmic entanglement growth there due to the absence of confinement, and is in perfect qualitative agreement with the conclusion of Ref.~\cite{Smith:2017aa}. 

\bibliography{MBLbib.bib}

\appendix

\section{Details on the numerical simulations}
  
In this section we provide a short description on the numerical investigation of the system dynamics described by Eq.~8 of the main text (MT). As stated before, we addressed this by employing a computationally optimized approach to the method of Krylov subspaces in order to avoid full diagonalization. In this scheme, we estimate the solutions to the time-dependent Schr\"odinger equation shown in Eq.~8MT by the polynomial approximation to $\ket{\Psi(t)}$ from within the Krylov subspace, given by

\begin{equation}\label{K_m}
\mathcal{K}_m = \textrm{span}\left\{\ket{\Psi_0}, H\ket{\Psi_0}, H^2\ket{\Psi_0}\dots,H^{m-1}\ket{\Psi_0}\right\}.
\end{equation} 
The optimal approximation is obtained by an Arnoldi decomposition procedure (equivalent to the Lanczos method for the case of Hermitian matrices) of the upper Hessenberg matrix $A_m$, defined as $A_m \equiv V^T_mHV_m$, where $V_m$ corresponds to the orthonormal basis resulting from the decomposition. $A_m$ can be seen as the projection of $H$ onto $\mathcal{K}_m$ with respect to the basis $V_m$. In the previous description $m$ is the dimension of the Krylov subspace.  

The desired solution is then approximated by
\begin{equation}\label{K_sol}
\ket{\Psi(t)}\approx V_{m}exp(-itA_m)\ket{e_1},
\end{equation}
where $\ket{e_1}$ corresponds to the first unit vector of the Krylov subspace. For a given dimension of the Hilbert space $\mathcal{D}$, the approximation shown in Eq.~\eqref{K_sol} becomes exact when $m \geq \mathcal{D}$, however, the method has been proven to be very effective even if $m \ll \mathcal{D}$ \cite{expokit}. For this particular approximated case, the much smaller matrix exponential can be evaluated using a Pad\`e approximation in conjunction to the well-known scaling-and-squaring algorithm \cite{brenes2017massively}. It has been proven theoretically \cite{expokiterror} that the error in this Krylov method behaves like $\mathcal{O}(e^{m-t||A||_{2}}(t||A||_{2}/m)^m)$ when $m \leq 2t||A||_{2}$, which indicates that the technique can be applied successfully if a time-stepping strategy is implemented along with error estimations. In practice, we estimate the intrinsic error of the algorithm by means of a forward-in-time evolution of an initial state; evaluating quantum observables for specific time steps and a selected charge configuration, then inverting the procedure with a backward-in-time evolution to determine that the numerical error stays within a desired bound. 

\section{Phase transition}

\begin{figure}[b]
\fontsize{24}{10}\selectfont
\centering
\includegraphics[width=1.01\columnwidth]{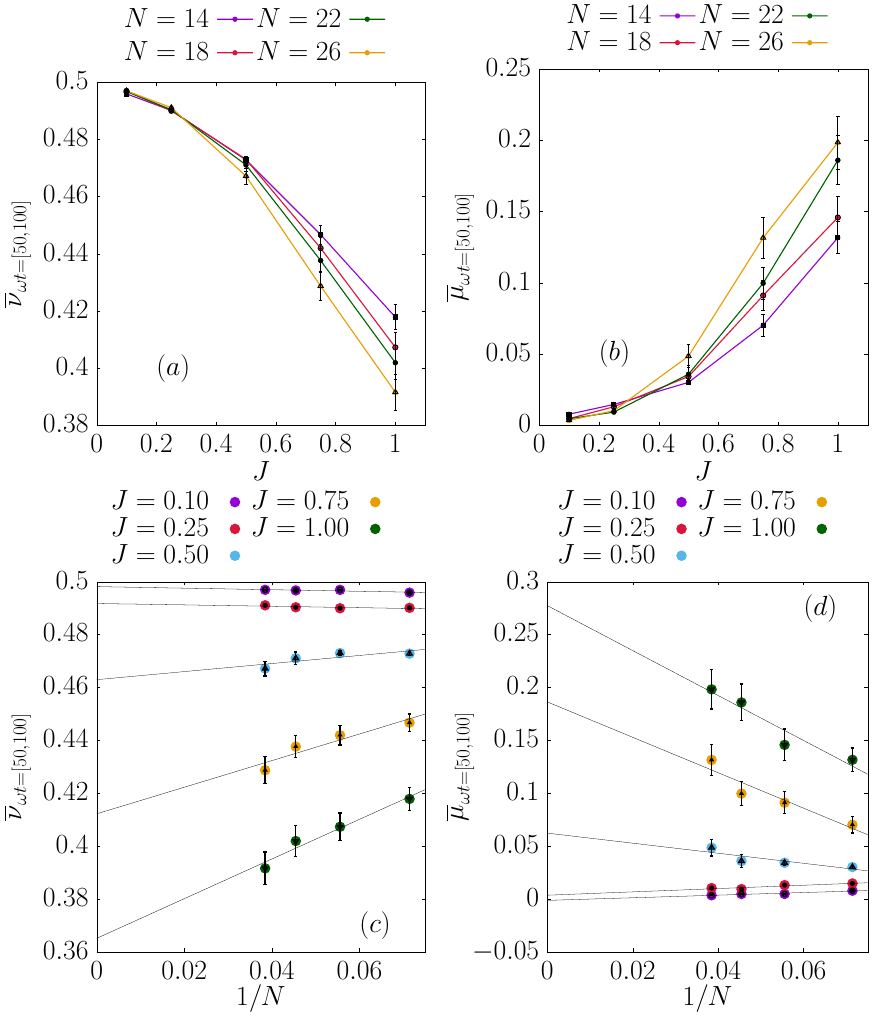}
\caption{(Color line) Estimating the location of the MBL transition as a function of the coupling strength $J$. (a), (b) Averaged long-time values of $\nu(t)$ and $\mu(t)$, respectively, as a function of $J$ for different system sizes $N$. (c), (d) Finite-size scaling analysis of memory effects in the system for different values of coupling strength. The absolute error extracted via the variance from a finite number of disorder realizations is shown as vertical bars for each point.}
\label{fig:5}
\end{figure} 

The contrasting behavior in the dynamics of the system from the weak to the strong coupling limit suggests the existence of an MBL transition. This behavior can be observed in panels (c) and (d) of Fig.~2 in the main text for the memory effects and in Fig.~3 of the main text for the spreading of quantum information. In this appendix we provide more details on the memory effects present in the system and give an estimate of the location  for the MBL transition as a function of the coupling strength. 

In Fig.~\ref{fig:5}a-b, we plot the limiting time values for both $\nu(t)$ and $\mu(t)$ for different values of coupling strength $J$. For the lowest shown simulated value of $J = 0.1$ the system exhibits ergodic dynamics as both of these observables reach their respective thermodynamic value. For $J = 0.25$, however, the dynamics already shows significant memory effects which appear to survive also in the thermodynamic, see the extrapolation In Fig.~\ref{fig:5}a-b we present the averaged long-time values of $\nu(t)$ and $\mu(t)$ (considered from $t = 50$ to $t = 100$ as described previously) as a function of $J$; in which for $J = 0.25$, both $\nu(t)$ and $\mu(t)$ show deviation from the values expected in the thermodynamic limit. The effect can be perceived more prominently by means of finite-size scaling, which we show in Fig.~\ref{fig:5}c-d. For every value of $J$ higher than 0.25 the dynamics show increasing memory effects with increasing system size, while the opposite happens for $J = 0.1$. Based on this observation we can approximate the value of the coupling strength parameter for which a phase transition from extended to MBL dynamics occurs in the range $0.1 \lesssim J_{critical} \lesssim 0.25$

\section{Evolution of lattice gauge theories and the role of superselection sectors}

In this appendix, we provide a more detailed discussion of the real-time dynamics of lattice gauge theories and the role of superselection sectors. We do so by first extending the treatment of the U(1) case discussed in the main text, and then, we provide an alternative interpretation of the results in Ref.~\cite{smith2017disorder}, where a $\mathbb{Z}_2$ quantum link model was shown to exhibit Anderson localization in the absence of disorder.

\paragraph{U(1) Wilson lattice gauge theory. -} The generic procedure in order to understand the role of superselection sectors in gauge invariant dynamics consists of three steps. 

{\it (i)} The first one is the choice of an initial state: we focus here on product states of the form 
\begin{equation}\label{PsiIn}
|\Psi\rangle_0 = |\Psi_f\rangle\otimes|\Psi_{g}\rangle
\end{equation}
where $|\Psi_f\rangle$ and $|\Psi_g\rangle$ are themselves also product states for the fermions and gauge fields, respectively. In particular, in the main text, we focus on a charge density like state for the fermions:
\begin{equation}
\label{PsiF}
|\Psi_f\rangle = |0101...\rangle
\end{equation}
and a homogeneous state for the gauge fields
\begin{equation}
\label{PsiG}
|\Psi_g\rangle = \bigotimes_{n=1}^{N-1} |\bar{L}_n\rangle\, , \, |\bar{L}_n\rangle = \frac{1}{\sqrt{3}} \left[ |-1\rangle_n + |0\rangle_n + |1\rangle_n \right].
\end{equation}
For all bonds the initial state represents an equal weight superposition of the three states $|-1\rangle_n, |0\rangle_n, |1\rangle_n$, with $L_n|a\rangle_n = a|a\rangle_n$. This is a starting configuration which leads to a modest disorder strength. Indeed, taking more generic initial states with gauge fields in arbitrary initial superpositions with unrestricted $a$'s generates an even stronger disorder landscape which is expected to lead to an even slower dynamics.

{\it (ii)} the second step is the decomposition of the initial state into different superselection sectors. A superselection sector is defined as a set of configurations $\{q_\alpha\}_{\alpha=1}^{N-1}$ such that the generators $G_n$ of the gauge transformation satisfy $G_n|\Psi_{\{q_\alpha\}}\rangle = q_n|\Psi_{\{q_\alpha\}}\rangle$. The values of $q_n$ correspond to the static charges of a given configuration at the site $n$, and is given by Eq. 5 of the main text. The vacuum sector corresponds to $q_n=0 \;\forall n$. The total number of superselection sectors is $3^{N-1}$ for our choice of initial state. 

Generically, each initial state of the form \eqref{PsiIn} can be decomposed into contributions from such different superselection sectors, see Eq. 5 of the main text. Let us give a simple example for a chain of 4 fermions and 3 gauge fields. In this case, the initial state can be composed into $3^{N-1}=9$ superselection sectors. For example, the state with $L_1=L_2=L_3=0$ has no static charges, while the state with $L_1=1, L_2=0, L_3=-1$ corresponds to having charges $q_1=1, q_2=-1, q_3 = -1, q_4 = 1$.

{\it (iii)} the third step is the evaluation of the effective dynamics within each superselection sector, since each of those is affected by a different charge distribution. 

In each sector, the corresponding Hamiltonian is made of two contributions, $H^{\{q_\alpha\}} = {H}_{\pm}+H_{In}^{\{q_\alpha\}}$. The first one describes electron-gauge coupling, which is not sensitive to background charges, and is given by:
\begin{eqnarray}
{H}_{\pm}\!&=&\!w\sum_{n=1}^{N-1}\left[{\sigma}_n^{+}{\sigma}_{n+1}^{-}+\textrm{h.c.}\right]
\end{eqnarray}
The second term originates from the electric field potential term, which is now a function of the fermionic populations only and reads:
\begin{eqnarray}
{H}_{In}\!&=&J\sum_{n=1}^{N-1} \left[ \frac{1}{2} \sum_{l=1}^n\left({\sigma}_l^z+q_\ell+(-1)^l\right) \right]^2.
\end{eqnarray}
This term can be conveniently decoupled into two parts, one containing two-body terms and one containing only one-body terms. The former results in:
\begin{eqnarray}
{H}_{ZZ}\!&=&\!\frac{J}{2}\sum_{n=1}^{N-2}\sum_{l=n+1}^{N-1}(N-l){\sigma}_n^{z} {\sigma}_l^z
\end{eqnarray}
which is related to the linear growth of Coulomb interactions in one-dimensional systems. 
The single spin terms become:
\begin{eqnarray}
{H}_{Z}^{\{q_\alpha\}}\!&=
&\!\frac{J}{2}\sum_{n=1}^{N-1} (\sum_{\ell=1}^n \sigma^z_\ell) \left[(\sum_{j=1}^n q_j) - n\text{mod}2 \right].
\label{eq:HamZ}
\end{eqnarray}
Crucially, this last part of the Hamiltonian depends explicitly on the superselection sector via $\{q_\ell\}$. This effectively generates an inhomogeneous disorder landscape. It is important to stress that this is a form of correlated disorder, since the charge configuration at each site $n$ depends on the static charges at the sites $m<n$.

We remark that this procedure is generic: gauge theories initialized in states which are product states of the form as in Eqs.~(\ref{PsiIn}-\ref{PsiG}) display real-time dynamics akin to disordered systems. This does not automatically lead to localization, whose emergence needs to be checked with non-perturbative methods, as we discuss in the main text. Moreover, it is worth noticing that there is a set of states which does not effectively feel disorder (for instance, due to homogeneous charge configurations). However, this represents a set of measure zero in the thermodynamic limit with contributions exponentially small in $N$.

\newpage

\paragraph{$\mathbb{Z}_2$ quantum link model. -} We now provide a discussion of the role of superselection sectors in the context of a $\mathbb{Z}_2$ quantum link model, where emergence of Anderson localization without disorder was recently reported~\cite{Smith:2017aa}.

In that case, $U_{n,n+1} = \sigma^z_n$, and $L_n = \sigma^x_n$. The authors employed a duality transformation (akin to the one used to demonstrate self-duality of the Ising model) to exactly map the dynamics of their QLM to free fermions evolving under random disorder configurations, depending on the initial state of their evolution. The same result can be obtained by a direct integration of the gauge fields~\cite{Hamer1997}, noting that, for $\mathbb{Z}_2$ theories, $L_{n+1} = L_n + \psi^\dagger_n\psi_n +q^{(2)}_n$, where $q^{(2)}_n=\pm1$ are $\mathbb{Z}_2$ static charges defined by the choice of the initial state. For example, let us consider initial states of the form $|\Psi\rangle_{CDW} = |0101...\rangle_\psi\otimes |\uparrow_z\uparrow_z...\rangle_\sigma$, that is, with fermions in a N\'eel state  and all gauge fields aligned along the $z$-axis, which can be decomposed in superselection sectors as:
\begin{equation}
|\Psi\rangle_{CDW} =\frac{1}{\mathcal{N}^{(2)}} \sum_{\{q_\alpha^{(2)}\}} |0101...\rangle_\psi\otimes |\bar{\sigma}^x_1\bar{\sigma}^x_2...\rangle_\sigma \, ,
\end{equation}
with $\bar{\sigma}^x_n = \bar{\sigma}^x_{n-1} + q^{(2)}_n + (-1)^n$, and $\mathcal{N}^{(2)}$ a normalization factor. In each of those sectors, the dynamics will be effectively disordered - the influence of random static charges will manifest in a random disordered potential with strength related to $J$.

\end{document}